\documentclass[aps,prd,preprint]{revtex4-1}
\usepackage {graphicx}
\usepackage[toc,page]{appendix}
\usepackage{amsmath}
\usepackage{tikz}
\newcommand{\be}{\begin{equation}}
\newcommand{\ee}{\end{equation}}
\newcommand{\bea}{\begin{eqnarray}}
\newcommand{\eea}{\end{eqnarray}}
\begin{document}
\title{Towards a spacelike characterization\\ of the null singularity inside a black hole}
\author{David Garfinkle}
\email{garfinkl@oakland.edu}
\affiliation{Dept. of Physics, Oakland University, Rochester, MI 48309, USA}
\affiliation{Leinweber Center for Theoretical Physics, Randall Laboratory of Physics, University of Michigan, Ann Arbor, MI 48109-1120, USA}

\date{\today}

\begin{abstract}

We propose a method for recognizing null singularities in a computer simulation that uses a foliation by spacelike surfaces.  The method involves harmonic time slicing as well as rescaled tetrad variables.  As a ``proof of concept'' we show that the method works in Reissner-Nordstrom spacetime. 

\end{abstract}


\maketitle

\section{Introduction}
The singularity theorems\cite{he}, starting with the work of Penrose in 1965\cite{penrose}, tell us that spacetime singularities occur in a wide range of circumstances.  However, these theorems tell us remarkably little about the nature of these singularities, saying only that some observer or light ray ends in finite affine parameter.  Since then, progress has come from a combination of analytic approximations, numerical simulations, and mathematical theorems using the methods of partial differential equations.  Remarkably, these results point to the presence of two different types of singularities: spacelike singularities and null singularities, both of which can be expected to occur in black hole interiors. 

The analytical approximations of Belinskii, Khalatnikov, and Lifschitz\cite{bkl} (collectively known as BKL) yield singularities that are spacelike, and local in the sense that as the singularity is approached time derivatives in the field equations become more important than space derivatives.  Support for the BKL picture of spacelike singularities has been provided by extensive sets of numerical simulations\cite{beverlyreview,allofus,dgharmonic,dgprl,ekpyrotic,meandfrans} going back to the work of Berger and Moncrief.\cite{beverlyandvince1,beverlyandvince2}  Additonal support is provided by the theorems of Fournodavlos, Rodnianski, and Speck\cite{rodnianskietal} on the behavior of the Einstein-scalar system.  Note that though the original BKL argument is local in the sense of not assuming any particular topology of space, this is not the case for the numerical work of\cite{beverlyreview}, or for the mathematical work of\cite{rodnianskietal}.  These results assume a spatial topology of $T^3$ and therefore do not directly apply to the sort of asymptotically flat gravitational collapse process that gives rise to black holes.

The evidence for null singularities begins with the fact that the inner horizon of the Kerr metric (and the Reissner-Nordstrom metric) is unstable.  Thus realistic gravitational collapse will not result in an inner horizon, but rather with that inner horizon being replaced by something else, likely something singular.  Remarkable analytic approximations, starting with the work of Poisson and Israel\cite{poissonisrael}, and then of Ori and Flanagan\cite{ori,oriandflanagan}, indicate that the result is indeed a singularity, but one that retains the inner horizon's property of being a null surface.  This picture received strong support from mathematical results on the stability of the Kerr metric due to Dafermos and Luk\cite{dafermosluk1,dafermosluk2} which show singular behavior on a surface that remains null.

Thus it is likely that the singularity inside a black hole has a piece that is spacelike and a piece that is null. One way that this might occur is as follows: the null singularity retains its character as a null surface, but only until those generators encounter a caustic, at which point the null singularity joins onto a spacelike singularity.

In order to see whether this picture is correct, it is necessary to do numerical simulations that are able to obtain the behavior of both parts of the singularity.  No numerical simulations are reported in this paper: instead we are concerned with possible methods to do the simulations and interpret the results.  

A simulation method typically uses a foliation by spacelike slices of constant coordinate time $T$.  In order to do a good job of simulating singularities, the time slices must approach arbitrarily close to the singularity as $T \to \infty$ without ever hitting it.  For spacelike singularities there are several different choices that have been shown to work, one of which is harmonic time.\cite{dgharmonic}  If the harmonic time slices also come abritrarily close to the null part of the singularity, then harmonic time slicing would be a suitable foliation for examining the singularities of black hole interiors.

Note that for a numerical simulation involving black hole formation from generic initial conditions, the foliation needs to (1) be compatible with a well posed initial value formulation of the Einstein field equations, (2) be compatible with asymptotic flatness, and (3) not depend on any symmetries for its construction.  Condition (3) is why we do not consider coordinate systems like Eddington-Finkelstein or Painleve-Gulstrand, which depend on spherical symmetry.  Condition (2) is why we do not consider foliation by constant mean curvature hypersurfaces, which has been used in simulations of singularities in the case of compact Cauchy surfaces ({\it i.e.} singularities in closed universes).  Harmonic time slicing satisfies all three conditions, and thus is worth exploring as a possibly suitable choice of foliation.

Since a harmonic time coordinate is a solution of the wave equation, it can be specified by giving its value and normal derivative on a Cauchy surface.  Since those initial conditions can be chosen differently, there is no uniqueness of harmonic time slicing of a given spacetime.  Nonetheless, the expectation that a solution of the wave equation is likely to blow up at the singularity leads us to expect that any one of a family of harmonic time coordinates might be suitable in the sense that the singularity is approached as the time coordinate goes to infinity.

However, even for a spacetime that is non-singular, when a foliation by spacelike surfaces approaches a null surface, the geometry of the spacelike surfaces becomes singular in the limit. 
More precisely: though the surface remains smooth in the limit, its intrinsic metric becomes degenerate and its extrinsic curvature blows up. Thus there is the issue of how to recognize a null singularity in a simulation using the usual foliation by spacelike surfaces.  To address this issue, we consider the scale invariant tetrad method of Uggla et al\cite{ugglaetal}, which in turn is based on the tetrad methods of Estabrook et al\cite{estabrook1,estabrook2}.  Scale invariant tetrad methods are often used in the treatment of spacelike singularities. In these methods the spacetime
is described in terms of a coordinate system ($T,{x^i}$) and a tetrad
(${{\bf e}_0},{{\bf e}_\alpha}$)  where ${{\bf e}_0}$ is orthogonal to the surfaces of constant $T$.  Here both the spatial coordinate
index $i$ and the spatial tetrad index $\alpha $ go from 1 to 3.  

The rescaling is done
using the extrinsic curvature $K_{ab}$ of the foliation as follows: first define the quantity $H$ by $H \equiv {{K^a}_a}/3$ (that is, $H$ is one third of the trace of the extrinsic curvature).  Now define the rescaled tetrad vectors ${\bf E}_0$ and ${\bf E}_\alpha$ by ${{\bf E}_0}\equiv{{\bf e}_0}/H$ and  
${{\bf E}_\alpha}\equiv{{\bf e}_\alpha}/H$.

A different rescaled tetrad method due to Ashtekar et al\cite{ashtekar} multiplies the tetrad vectors by $\sqrt h$ (where $h$ is the determinant of the spatial metric) rather than dividing by $H$.  These rescaled tetrad vectors are then tensor densities, which allows the system of evolution equations to be treated using Hamiltonian methods.  The Uggla et al rescaled tetrad vectors are not tensor densities, but they are nonetheless perfectly good (foliation dependent) geometric quantities since they are constructed from the tetrad and the extrinsic curvature of the foliation surfaces. 

As the singularity is approached ${{\bf e}_0}$ and ${{\bf e}_\alpha}$ blow up, but so does $H$, and in such a way that ${\bf E}_0$ and ${\bf E}_\alpha$ are finite.  Furthermore, for spacelike singularities described by the BKL conjecture, the spacelike rescaled vectors ${\bf E}_\alpha$ vanish as the singularity is approached.  Since all spatial derivative terms in the field equations occur multiplied by ${\bf E}_\alpha$, this property that ${\bf E}_\alpha$ vanishes at the singularity gives a concrete version of the BKL statement that at the singularity spatial derivatives in the field equations become negligible compared to time derivatives. (As two examples of this sort of behavior, the appendix shows the behavior of the rescaled tetrad in the Schwarzschild spacetime and in the Kasner spacetime, for a foliation by constant harmonic time hypersurfaces approaching the singularity).

As a ``proof of concept'' for these ideas, this paper involves a demonstration that they work in the Reissner-Nordstrom spacetime.  That is we will do the following: (1) find a harmonic time coordinate $T$ whose level sets approach arbitrarily close to the inner horizon of Reissner-Nordstrom as $T \to \infty$.  (2) compute the rescaled tetrad vectors appropriate to this foliation, and (3) show that the rescaled tetrad vectors have a finite limit as $T \to \infty$.

\section{Reissner-Nordstrom}
\label{RN}
We begin with some well known properties of the Reissner-Nordstrom spacetime.  The metric in the usual $(t,r)$ coordinate system takes the form
\be
d{s^2} = - F d{t^2} + {F^{-1}} d {r^2} + {r^2} (d {\theta^2} + {\sin^2}\theta d {\phi^2}) \; \; \; ,
\label{rn}
\ee
where 
\be
F = 1 - {\frac {2M} r} + {\frac {Q^2} {r^2}} \; \; \; ,
\label{rnF}
\ee
and $M$ and $Q$ are respectively the mass and charge of the black hole.
The outer and inner horizons are at $r={r_\pm}$ where ${r_\pm}=M \pm {\sqrt {{M^2}-{Q^2}}}$.  Expressing $F$ in terms of $r_+$ and $r_-$ we obtain
\be
F = {\frac 1 {r^2}} (r-{r_+})(r-{r_-}) \; \; \; .
\label{rnF2}
\ee
The null coordinates $v$ and $u$ are defined by
\bea
dv = dt + {F^{-1}} d r \; \; \; ,
\\
du = dt - {F^{-1}} d r \; \; \; .
\eea
This leads to the following expressions for the metric in $(v,r)$ and $(u,r)$ coordinates 
\bea
d{s^2} = - F d{v^2} + 2 d v d r + {r^2} (d {\theta^2} + {\sin^2}\theta d {\phi^2}) \; \; \; ,
\\
d{s^2} = - F d{u^2} - 2 d u d r + {r^2} (d {\theta^2} + {\sin^2}\theta d {\phi^2}) \; \; \; .
\eea
Most of the spacetime can be coverered by a set of overlapping $(v,r)$ and $(u,r)$ coordinate patches, and for our purposes that will be sufficient. For a more complete set of Kruskal type coordinates, see Brill and Graves\cite{brill} and Hawking and Ellis.\cite{he}

\section{harmonic time}

We want a harmonic time coordinate $T$ where the surfaces of constant $T$ extend inside the black hole, and therefore we need for $T$ to not blow up at $r={r_+}$.  We try the ansatz $T=v + h(r)$ and obtain from ${\nabla ^a}{\nabla_a}T=0$ the following equation for $h$
\be
0 = {\frac d {dr}}\left [ {r^2} \left ( 1 + F {\frac {dh} {dr}} \right ) \right ] \; \; \; .
\ee
Thus there is a constant $c$ for which 
\be
c - {r^2} = (r-{r_+})(r-{r_-}) {\frac {dh} {dr}} \; \; \; .
\ee
In order for $h$ to not blow up at $r={r_+}$ we must choose $c={r_+^2}$ which leads to 
\be
{\frac {dh} {dr}} = -1 \; {\frac {r+{r_+}} {r-{r_-}}} \; \; \; ,
\ee
and thus
\be
h = - r - 2 M \ln \left ( {\frac r {r_-}} - 1 \right ) \; \; \; ,
\ee
which in turn yields
\be
T = v - r - 2 M \ln \left ( {\frac r {r_-}} - 1 \right ) \; \; \; .
\label{harmonicT}
\ee

\section{tetrad and rescaled tetrad}

With $T$ as our time slicing coordinate, the tetrad vector ${e_{0a}}$ is the unit vector in the direction of ${\nabla_a} T$.  We have 
\be
{\nabla _a}T = {\nabla _a} v - {\frac {r+{r_+}} {r-{r_-}}} {\nabla _a} r \; \; \; .
\ee
However we want to know the behavior of the tetrad near the inner horizon, which is the limit as $r\to {r_-}$ at constant $u$.  We therefore want to express all quantities in a $(u,r)$ coordinate system in the region ${r_-} < r < {r_+}$.  Since $dv = du + 2 {F^{-1}}dr$, some straightforward algebra leads to 
\be
{\nabla _a} T = {\nabla_a} u + {F^{-1}} \left ( 1 + {\frac {r_+^2} {r^2}} \right ) {\nabla _a} r \; \; \; .
\label{Tur}
\ee
Integrating eqn. (\ref{Tur}) we obtain
\be
T = u + r + {\frac 1 {{r_+}-{r_-}}} \left [ 2 {r_+ ^2} \ln \left ( 1 - {\frac r {r_+}} \right ) \; - \; ({r_+ ^2}+{r_- ^2}) \ln \left ( {\frac r {r_-}} - 1 \right ) \right ] \; \; \; ,
\ee
which yields 
\be
{\frac r {r_-}} - 1 = {{\left ( 1 - {\frac r {r_+}} \right ) }^{-2{r_+ ^2}/({r_+ ^2} + {r_- ^2})}} \; \exp \left [ \left ( {\frac {{r_+}-{r_-}} {{r_+ ^2} + {r_- ^2}}} \right ) (u + r - T) \right ] \; \; \; .
\ee
Thus we see that at fixed $u$ we have $r \to {r_-}$ as $T \to \infty$.

Raising the index of eqn. (\ref{Tur}) we obtain
\be
{\nabla ^a}T = {\frac {r_+^2} {r^2}} {{\left ( {\frac \partial {\partial r}} \right ) }^a}
- {F^{-1}} \left ( 1 + {\frac {r_+^2} {r^2}} \right ) {{\left ( {\frac \partial {\partial u}} \right ) }^a} \; \; \; .
\ee
Since ${e_0}^a$ is a unit vector in the direction of ${\nabla^a}T$ we obtain 
\be
{{e_0}^a} = {e^{-W/2}} \left [ {\frac {r_+^2} {r^2}} {{\left ( {\frac \partial {\partial r}} \right ) }^a}
- {F^{-1}} \left ( 1 + {\frac {r_+^2} {r^2}} \right ) {{\left ( {\frac \partial {\partial u}} \right ) }^a} \right ] \; \; \; ,
\ee
where the quantity $W$ is defined by
\be
W \equiv \ln \left [ {\frac {r+{r_+}} {r-{r_-}}} \left ( 1 + {\frac {r_+^2} {r^2}} \right ) \right ] \; \; \; .
\label{Wdef}
\ee
For the sign of ${e_0}^a$ we use the convention of \cite{ugglaetal} that ${e_0}^a$ points away from the singularity.

We choose the tetrad vector ${e_1}^a$ to be the unit vector in the $(t,r)$ plane orthogonal to ${e_0}^a$ which results in the expression
\be
{{e_1}^a} = {e^{-W/2}} \left [  {{\left ( {\frac \partial {\partial r}} \right ) }^a}
- {F^{-1}} \left ( 1 + {\frac {r_+^2} {r^2}} \right ) {{\left ( {\frac \partial {\partial u}} \right ) }^a} \right ] \; \; \; .
\ee
To complete the tetrad we need two orthonormal vectors in the two-sphere direction, which we choose to be
\bea
{{e_2}^a} = {\frac 1 r} {{\left ( {\frac \partial {\partial \theta}} \right ) }^a} \; \; \; ,
\\
{{e_3}^a} = {\frac 1 {r\sin \theta}} {{\left ( {\frac \partial {\partial \phi}} \right ) }^a}
\; \; \; .
\eea

The rescaled tetrad is given by 
$({{E_0}^a},{{E_1}^a},{{E_2}^a},{{E_3}^a})= ({{e_0}^a},{{e_1}^a},{{e_2}^a},{{e_3}^a})/H$ where $H$ is given by $H =(1/3) {\nabla _a}{{e_0}^a}$.  Since ${\sqrt {-g}}={r^2} \sin \theta$ we have
\be
H = {\frac 1 {3 {r^2} \sin \theta}} {\partial _\mu} ({r^2} \sin \theta {{e_0}^\mu}) 
= - {\frac 1 6} {\frac {r_+^2} {r^2}} {e^{-W/2}} {\frac {dW} {dr}} \; \; \; .
\ee 
This in turn yields the following expression for the rescaled tetrad
\bea
{{E_0}^a} &=& {{\left ( {\frac M {r+{r_+}}} + {\frac {{r_+^2}(r-{r_-})} {r ({r^2} + {r_+^2})}} \right ) }^{-1}} \left [ 3 (r-{r_-}) {{\left ( {\frac \partial {\partial r}} \right ) }^a}  - {\frac {3{r^2}} {r-{r_+}}} \left ( {\frac {r^2} {r_+^2}} + 1 \right ) {{\left ( {\frac \partial {\partial u}} \right ) }^a} \right ] 
\\
{{E_1}^a} &=& {{\left ( {\frac M {r+{r_+}}} + {\frac {{r_+^2}(r-{r_-})} {r ({r^2} + {r_+^2})}} \right ) }^{-1}} \left [ 3 (r-{r_-}) {\frac {r^2} {r_+^2}}{{\left ( {\frac \partial {\partial r}} \right ) }^a} 
 - {\frac {3{r^2}} {r-{r_+}}} \left ( {\frac {r^2} {r_+^2}} + 1 \right ) {{\left ( {\frac \partial {\partial u}} \right ) }^a} \right ] 
\\
{{E_2}^a} &=& {{\left ( {\frac M {r+{r_+}}} + {\frac {{r_+^2}(r-{r_-})} {r ({r^2} + {r_+^2})}} \right ) }^{-1}} {\frac 3 {r_+^2}} {\sqrt {(r-{r_-})(r+{r_+})({r^2} + {r_+^2})}} {{\left ( {\frac \partial {\partial \theta}} \right ) }^a}
\\
{{E_3}^a} &=& {{\left ( {\frac M {r+{r_+}}} + {\frac {{r_+^2}(r-{r_-})} {r ({r^2} + {r_+^2})}} \right ) }^{-1}} {\frac 3 {{r_+^2}\sin \theta}} {\sqrt {(r-{r_-})(r+{r_+})({r^2} + {r_+^2})}} {{\left ( {\frac \partial {\partial \phi}} \right ) }^a}
\eea

In the limit as $r\to {r_-}$ we have
\bea
{{E_0}^a} &\to& {\frac {6 {r_-^2}({r_+^2}+{r_-^2})} {{r_+^2}({r_+}-{r_-})}} {{\left ( {\frac \partial {\partial u}} \right ) }^a} \; \; \; ,
\\
{{E_1}^a} &\to& {\frac {6 {r_-^2}({r_+^2}+{r_-^2})} {{r_+^2}({r_+}-{r_-})}} {{\left ( {\frac \partial {\partial u}} \right ) }^a} \; \; \; ,
\\
{{E_2}^a} &\to& 0 \; \; \; ,
\\
{{E_3}^a} &\to& 0 \; \; \; .
\eea
Note that the limiting values of ${{E_0}^a}$ and ${{E_1}^a}$ are colinear, and that both point in the $u$ direction, which is the direction of the null generators of the horizon.  Though, our calculation is performed in a particular coordinate system, the statements about colinearity and coinciding with the null generators of the horizon are geometric statements. 
\section{Conclusion}
\label{Conclude}
We have shown that the rescaled tetrad has a nonsingular limit as the inner horizon is approached.  In this limit, two rescaled tetrad vectors are nonzero.  Furtheremore these vectors are colinear and both point along the null generators of the inner horizon.  This behavior is very different from what happens at spacelike singularities, where only ${E_0}^a$ has a nonzero limit.  Thus we have a way of recognizing the difference between spacelike singualrities and null singularities.  

More precisely, we have a proposed method for how to simulate the formation of black holes, including the singularities that form in the black hole interior. This method includes using a foliation by harmonic time slices, a set of tetrad variables, and an examination of the rescaled tetrad to see which parts of the singularity are spacelike and which parts are null.

The next (and far more difficult) step is to perform simulations of gravitational collapse and black hole formation using this proposed method.  Such simulations are work in progress.

\section*{Acknowledgments}

It is a pleasure to thank Frans Pretorius, Amos Ori, Mihalis Dafermos, Eric Poisson, and Abhay Ashtekar for helpful discussions.  This work was supported by NSF Grant PHY-2102914. 

\appendix

\section{Schwarzschild and Kasner singularity in rescaled tetrad variables}
To get an idea of how the rescaled tetrad variables behave in the case of spacelike singularities described by the BKL conjecture, we consider two examples: the Schwarzschild spacetime and the Kasner spacetime.  

In the Schwarzschild spacetime, the line element is given by 
\be
d{s^2} = - \left ( 1 - {\frac {2M} r} \right ) d {t^2} + {{ \left ( 1 - {\frac {2M} r} \right ) }^{-1}} d {r^2} + {r^2} (d{\theta ^2} + {\sin ^2} \theta d {\phi ^2} ) \; \; \; .
\label{schwarzschild}
\ee

We treat the region inside the horizon and use a foliation by surfaces of constant harmonic time $T$.  Since $r$ is a timelike coordinate inside the horizon, we try the ansatz $T=T(r)$.  
The equation ${\nabla _a}{\nabla ^a}T=0$ then becomes
\be
{\frac d {dr}} \left ( r(2M-r) {\frac {dT} {dr}} \right ) = 0 \; \; \; .
\ee
The solution is 
\be 
T = {c_0} \ln \left ( {\frac {2M} r} - 1 \right ) \; \; \; ,
\ee
for some constant $c_0$ (where without loss of generality we have set the other integration constant to zero).  We choose ${c_0}=1$.  Solving for $r$ we obtain
\be
r = 2 M {{({e^T} + 1)}^{-1}} \; \; \; .
\ee
Note that the singularity at $r=0$ is approached as $T \to \infty$.  

Since the usual Schwarzschild coordinate $t$ is spacelike inside the horizon, we introduce the coordinate $\rho$ by $\rho = t/(2M)$.  In the $(T,\rho,\theta,\phi)$ coordinate system the line element becomes
\be
d {s^2} = 4 {M^2} \left [ - {e^T}{{({e^T}+1)}^{-4}} d {T^2} + {e^T} d {\rho ^2} + {{({e^T}+1)}^{-2}} ( d {\theta ^2} + {\sin ^2} \theta d {\phi ^2} ) \right ] \; \; \; .
\ee
The vector ${{\bf e}_0}^a$ is the unit vector orthogonal to the surfaces of constant $T$ and pointing away from the singularity, so we find
\be
{{{\bf e}_0}^a} = {\frac {-1} {2M}} {e^{-T/2}} {{({e^T}+1)}^2} {{\left ( {\frac \partial {\partial T}} \right ) }^a} \; \; \; .
\label{sche0}
\ee
A convenient choice for the spatial triad is 
\bea
{{{\bf e}_1}^a} &=& {\frac 1 {2M}} {e^{-T/2}} {{\left ( {\frac \partial {\partial \rho}} \right ) }^a} \; \; \; ,
\\
{{{\bf e}_2}^a} &=& {\frac 1 {2M}} ({e^T}+1) {{\left ( {\frac \partial {\partial \theta}} \right ) }^a} \; \; \; ,
\\
{{{\bf e}_3}^a} &=& {\frac 1 {2M}} ({e^T}+1) {\frac 1 {\sin \theta}} {{\left ( {\frac \partial {\partial \phi}} \right ) }^a} \; \; \; .
\eea

The quantity $H$ is given by $H=(1/3){\nabla _a}{{{\bf e}_0}^a}$, from which we find
\be
H = {\frac 1 {3{\sqrt g}}} {\frac \partial {\partial T}} ({\sqrt g} {{{\bf e}_0}^T}) = 
{\frac 1 {12 M}}{e^{-T/2}}({e^T}+1)(3{e^T}-1) \; \; \; .
\ee
The rescaled tetrad is then given by 
\bea
{{{\bf E}_0}^a} &=& {\frac {-6 ({e^T}+1)} {3{e^T}-1}} {{\left ( {\frac \partial {\partial T}} \right ) }^a} \; \; \; ,
\\
{{{\bf E}_1}^a} &=& {\frac 6 {({e^T}+1)(3{e^T}-1)}} {{\left ( {\frac \partial {\partial \rho}} \right ) }^a} \; \; \; ,
\\
{{{\bf E}_2}^a} &=& {\frac {6 {e^{T/2}}} {3{e^T}-1}} {{\left ( {\frac \partial {\partial \theta}} \right ) }^a} \; \; \; ,
\\
{{{\bf E}_3}^a} &=& {\frac {6 {e^{T/2}}} {(3{e^T}-1)\sin \theta}} {{\left ( {\frac \partial {\partial \phi}} \right ) }^a} \; \; \; .
\eea

As the singularity is approached, that is in the limit as $T \to \infty$ we have
\bea
{{{\bf E}_0}^a} &\to & -2 {{\left ( {\frac \partial {\partial T}} \right ) }^a} \; \; \; ,
\\
{{{\bf E}_1}^a} & \to & 0 \; \; \; ,
\\
{{{\bf E}_2}^a} &\to & 0 \; \; \; ,
\\
{{{\bf E}_3}^a} &\to & 0 \; \; \; .
\eea
That is, the timelike rescaled tetrad vector approaches a nonzero quantity, while the rescaled spatial triad vectors vanish.  This is the normal behavior for singularities in the BKL picture.

We now turn to the Kasner spacetime, a homogeneous, but anisotropic, solution of the vacuum Einstein equation.  The line element is given by 
\be
d {s^2} = - d{t^2} + {t^{2{p_1}}} d {x^2} + {t^{2{p_2}}} d {y^2} + {t^{2{p_3}}} d {z^2} \; \; \; .
\label{Kasner}
\ee
Here the $p_i$ are constants satisfying 
\bea
{\sum _i} \; {p_i}&=&1 \; \; \; ,
\label{pcondition1}
\\
{\sum _i}\; {p_i ^2} &=& 1 \; \; \; .
\label{pcondition2}
\eea
Note that it follows from eqn. (\ref{pcondition1}) that ${\sqrt g} = t$.  

We want a foliation by surfaces of constant harmonic time $T$.  We try the ansatz $T=T(t)$.  
The equation ${\nabla _a}{\nabla ^a}T=0$ then becomes
\be
{\frac d {dt}} \left ( t{\frac {dT} {dt}} \right ) = 0 \; \; \; .
\ee
The solution is 
\be 
T = {c_0} \ln t \; \; \; ,
\ee
for some constant $c_0$  (where without loss of generality we have set the other integration constant to zero).  The singularity is at $t=0$.  In order that the singularity is approached as $T \to \infty$ we must choose $c_0$ to be negative.  We choose ${c_0}=-1$.  Solving for $t$ we obtain
\be
t= {e^{-T}} \; \; \; .
\ee
Thus the Kasner line element of eqn. (\ref{Kasner}) takes the form
\be
d {s^2} = - {e^{- 2 T}}d{T^2} + {e^{-2{p_1}T}} d {x^2} +  {e^{-2{p_1}T}} d {y^2} + 
{e^{-2{p_1}T}} d {z^2} \; \; \; .
\label{Kasner2}
\ee
Note that in this coordinate system it follows from eqn. (\ref{pcondition1}) that ${\sqrt g} = {e^{-2T}}$.   

The vector ${{\bf e}_0}^a$ is the unit vector orthogonal to the surfaces of constant $T$ and pointing away from the singularity, so we find
\be
{{{\bf e}_0}^a} = - {e^T} {{\left ( {\frac \partial {\partial T}} \right ) }^a} \; \; \; .
\label{kasnere0}
\ee
A convenient choice for the spatial triad is 
\bea
{{{\bf e}_1}^a} &=& {e^{{p_1}T}} {{\left ( {\frac \partial {\partial x}} \right ) }^a} \; \; \; ,
\\
{{{\bf e}_2}^a} &=& {e^{{p_2}T}} {{\left ( {\frac \partial {\partial y}} \right ) }^a} \; \; \; ,
\\
{{{\bf e}_3}^a} &=& {e^{{p_3}T}} {{\left ( {\frac \partial {\partial z}} \right ) }^a} \; \; \; .
\eea

The quantity $H$ is given by $H=(1/3){\nabla _a}{{{\bf e}_0}^a}$, from which we find
\be
H = {\frac 1 {3{\sqrt g}}} {\frac \partial {\partial T}} ({\sqrt g} {{{\bf e}_0}^T}) = 
{\frac 1 3}{e^T} \; \; \; .
\ee
The rescaled tetrad is then given by 
\bea
{{{\bf E}_0}^a} &=& -3 {{\left ( {\frac \partial {\partial T}} \right ) }^a} \; \; \; ,
\\
{{{\bf E}_1}^a} &=& 3 {e^{({p_1}-1)T}} {{\left ( {\frac \partial {\partial x}} \right ) }^a} \; \; \; ,
\\
{{{\bf E}_2}^a} &=& 3 {e^{({p_2}-1)T}} {{\left ( {\frac \partial {\partial y}} \right ) }^a} \; \; \; ,
\\
{{{\bf E}_3}^a} &=& 3 {e^{({p_3}-1)T}} {{\left ( {\frac \partial {\partial z}} \right ) }^a} \; \; \; .
\eea

Note that it follows from eqn. (\ref{pcondition2}) that ${p_i} \le 1$ for each $p_i$ with equality only in the case where one of the $p_i$ is 1 and the other two are 0.  This later case is not a singular spacetime, but is instead Minkowski spacetime in an unusual coordinate system.  Thus in our consideration of singular Kasner spacetimes we have ${p_i} < 1$.  It then follows that 
as the singularity is approached, that is in the limit as $T \to \infty$ we have
\bea
{{{\bf E}_0}^a} &\to & -3 {{\left ( {\frac \partial {\partial T}} \right ) }^a} \; \; \; ,
\\
{{{\bf E}_1}^a} & \to & 0 \; \; \; ,
\\
{{{\bf E}_2}^a} &\to & 0 \; \; \; ,
\\
{{{\bf E}_3}^a} &\to & 0 \; \; \; .
\eea
That is, the timelike rescaled tetrad vector approaches a nonzero quantity, while the rescaled spatial triad vectors vanish.  This is the normal behavior for singularities in the BKL picture.

\end{document}